\documentclass[final]{raa06}           %  For uploading to arXiv and final 
\usepackage{graphicx,times}             %for PS/EPS graphics inclusion, new
\usepackage{natbib}
\usepackage{amssymb,amsmath}
\bibpunct{(}{)}{;}{a}{}{,}
\usepackage[colorlinks=true, citecolor=blue]{hyperref}%
\usepackage{graphicx,kantlipsum,setspace}
\usepackage{caption}
\usepackage{graphics,epsf}
\usepackage{amsmath}                % American Mathematical Society package
\usepackage{amsfonts}               % American Mathematical Society fonts
\usepackage{amssymb}                % American Mathematical Society symbol
\usepackage{epsfig}                 % EPS figures
\usepackage{appendix}
\usepackage{graphicx}
\usepackage{float}
\usepackage{color}
\usepackage{multirow}
\usepackage{colortbl}
\usepackage[para,online,flushleft]{threeparttable}
\usepackage{xcolor}

\hypersetup{citecolor=blue, % color for \cite{...} links
            linkcolor=red, % color for \ref{...} links
            menucolor=blue, % color for Acrobat menu buttons
            urlcolor=blue}  % color for \url{...} links
\newcommand{\yr}{{~\rm yr}}

\begin{document}

   \title{The depletion of the red supergiant envelope radiative zone during common envelope evolution
%\footnotetext{$*$ Supported by the National Natural Science Foundation of China.}
}
%   \subtitle{I. Place Your Subtitle Here}

   \volnopage{Vol.0 (20xx) No.0, 000--000}      %%preserved for Editor. DOn't remove!
   \setcounter{page}{1}          %%starting page, preserved for Editor. DOn't remove!

   \author{Tamar Cohen %\orcidlink{0009-0001-0720-6816}
      \inst{1}
 \and Noam Soker % \orcidlink{0000-0003-0375-8987} 
 \inst{1}
   }

   \institute{Department of Physics, Technion, Haifa, 3200003, Israel;   {\it  tamarco@campus.technion.ac.il; soker@physics.technion.ac.il}\\
%        \and
%             Full institute address for the third author\\
\vs\no
   {\small Received~~20xx month day; accepted~~20xx~~month day}}

\abstract{
We conduct one-dimensional stellar evolution simulations of red supergiant (RSG) stars that mimic common envelope evolution (CEE) and find that the inner boundary of the envelope convective zone moves into the initial envelope radiative zone. The envelope convection practically disappears only when the RSG radius decreases by about an order of magnitude or more.  The implication is that one cannot split the CEE into one stage during which the companion spirals-in inside the envelope convective zone and removes it, and a second slower phase when the companion orbits the initial envelope radiative zone and a stable mass transfer takes place. At best, this might take place when the orbital separation is about several solar radii. However, by that time other processes become important. We conclude that as of yet, the commonly used alpha-formalism that is based on energy considerations is the best phenomenological formalism.
}

 \authorrunning{T. Cohen and N. Soker}            
\titlerunning{Depletion of the RSG radiative zone during CEE}  
   
      \maketitle

% ==========================================================
\section{INTRODUCTION}
\label{sec:intro}
% ==========================================================

Unsolved questions regarding the common envelope evolution (CEE) have stimulated different types of theoretical studies in recent years. One type includes studies of three-dimensional hydrodynamical CEE simulations (e.g., \citealt{TaamRicker2010, DeMarcoetal2011, Passyetal2012, RickerTaam2012, Nandezetal2014, Staffetal2016MN8, Kuruwitaetal2016, Ohlmannetal2016a, Ohlmannetal2016b, DeMarcoIzzard2017, Galavizetal2017, Iaconietal2017b, LawSmithetal2020, GlanzPerets2021a, GlanzPerets2021b, GonzalezBolivar2022, Lauetal2022a, Lauetal2022b, Morenoetal2022, GagnierPejcha2023}; for a recent thorough review with more references see \citealt{RoepkeDeMarco2023}; for a relativistic study in 2D see \citealt{CruzOsorioRezzolla2020}). Most three-dimensional CEE simulations that do not include accretion energy nor recombination energy encounter the problems that they do not manage to eject the entire envelope and/or that the final orbital separation of the core-companion binary system is larger than what observations show (e.g., \citealt{IaconiDeMarco2019}). 

The second type of studies examines extra energy sources to remove the envelope. Some studies consider the recombination energy of helium, and possibly also of hydrogen, of the ejected common envelope (e.g., \citealt{Nandezetal2015, IvanovaNandez2016, Kruckowetal2016, WangC2016}, and some 3D simulations cited above). However, some other studies suggest that convection in the envelope of giants is very efficient in carrying the recombination energy close to the photosphere where the extra energy rapidly diffuses out (e.g., \citealt{Sabachetal2017, Gricheneretal2018, WilsonNordhaus2019, WilsonNordhaus2020, WilsonNordhaus2022}). We take the view that most of the recombination energy is radiated away, giving rise to a transient event during the CEE, at least during early times of several to several tens times the dynamical time of the system. We therefore neglect the contribution of recombination to the energy budget of unbinding the envelope.

Another possible extra energy source is the accretion of envelope mass onto the compact companion. This energy is carried to the envelope via jets (e.g., \citealt{ArmitageLivio2000, Chevalier2012} for neutron star companions, \cite{SchreierSoker2016, Shiberetal2016} for main sequence companions, and \citealt{Soker2016Rev,Soker2022Rev} for general reviews). Some recent simulations include jets in three-dimensional CEE or grazing envelope evolution simulations (e.g., \citealt{MorenoMendezetal2017, ShiberSoker2018, LopezCamaraetal2019, Schreieretal2019, Shiberetal2019, LopezCamaraetal2020MN, Hilleletal2022, LopezCamaraetal2022, Zouetal2022, Schreieretal2023}). 

A third type of theoretical studies examines new prescriptions to estimate the final orbital separation $a_f$ (see also \citealt{Tranietal2022} for studying the eccentricity). These studies aim at replacing the commonly used, e.g., in many population synthesis studies (e.g., \citealt{Grichener2023, Huetal2023, Zhuetal2023} for some recent studies), alpha-prescription that assumes that a fraction $\alpha_{\rm CE}$ of the released orbital energy, $\Delta E_{\rm orb}$, goes to unbind the envelope, which has an initial binding energy of $E_{\rm bind,0}$ (from old studies, e.g., \citealt{vandenHeuvel1976},  \citealt{TutukovYungelson1979} who called the parameter $\beta$, and \citealt{Webbink1984} to newer studies, e.g.,  \citealt{WuDetal2019, Wuetal2020, Geetal2022}). Namely, $\alpha_{\rm CE} \Delta E_{\rm orb} = E_{\rm bind,0}$. In a recent study \cite{DiStefanoetal2023} propose a phenomenological prescription that is based on angular momentum conservation. An earlier prescription that is based on angular momentum is the $\gamma$-formalism \citep{NelemansTout2005}, that has been used in some population synthesis studies (e.g., \citealt{Toonenetal2012}).  
Prescriptions based on angular momentum conservation to estimate the final orbital separation suffer from the problem that because the ratio of the initial angular momentum to the final orbital angular momentum is very large, a small change in model parameters, which results from uncertainties at the initial CEE phases, will lead to large variations in the values of the final orbital separation $a_f$ (e.g., \citealt{Webbink2008} and section 5.2.2 of \citealt{Ivanovaetal2013}). The specific angular momentum prescription that \cite{DiStefanoetal2023} propose has another severe drawback. It predicts that very low mass companions are able to remove the envelope of giants, i.e., their results suggest that a companion of one percent of the giant's core mass can remove the entire envelope without much decrease, or even increase, in the orbital separation. Extrapolating their results we find that Mercury will be able to eject the envelope of the Sun along the red giant branch of the Sun and survive. 

In another new paper \cite{HiraiMandel2022} propose a two-stage formalism for CEE of massive stars. In the first stage they propose that the compact companion rapidly spirals-in inside the outer convective zone of a red supergiant (RSG) envelope. For that they propose to use the $\alpha$-formalism. In the second phase, they propose, the companion evolves on a long thermal time scale with a stable mass transfer from the inner radiative zone of the envelope. Using a one-dimensional code (section \ref{sec:Numerics}) we evolve RSG models (sections \ref{sec:RadiativeZone} and \ref{sec:RadiativeZonedynamic}) and show that their assumption regarding the unchanged structure of the radiative zone does not hold (section \ref{sec:Implications}). 
We summarize our short study in section \ref{sec:Summary} where we also present our view on the best CEE formalism.   

% =============================================
\section{Numerical procedure}
\label{sec:Numerics}
% =============================================

Our goal is to follow the evolution of the radiative zone as the RSG star loses mass at a high rate during the CEE. For that we evolve stellar models with a zero age main sequence (ZAMS) mass of $M_{\rm ZAMS}=12 M_\odot$ and with initial metallicities of $Z=0.001$,   $Z=0.01$, or $Z=0.02$. We use version r22.05.1 of \textsc{mesa} (Modules for Experiments in Stellar Astrophysics; \citealt{Paxtonetal2011, Paxtonetal2013, Paxtonetal2015, Paxtonetal2018, Paxtonetal2019}). 
The opacities, which are important for the location of the radiative and convective zones, are primarily from OPAL \citep{IglesiasRogers1996}, with low-temperature data from \citet{Fergusonetal2005} and the high-temperature, Compton-scattering dominated regime by \citet{Poutanen2017}. We turn off wind mass loss as this is not important for our study. 

When the star becomes a RSG we stop the single-star evolution, i.e., we change the mass-loss-rate parameter of a single star evolution in  \textsc{mesa} and substantially increase the mass loss rate to mimic CEE. We manually remove mass from the RSG envelope at a high rate (much above that of a regular wind) of $\dot M_{\rm CEE} = 0.01 M_\odot \yr^{-1}$ (section \ref{sec:RadiativeZone}) or at a very-high rate of $\dot M_{\rm CEE,d} = 3 M_\odot \yr^{-1}$ (section \ref{sec:RadiativeZonedynamic}). Mass is lost from the outer layer of the star as the default setting in \textsc{MESA} and according to the expectation of CEE, but does not include non-spherical effects that must take place in CEE. 
This mass removal mimics the envelope ejection process during the CEE. The high mass removal rate removes the envelope on a thermal timescale of hundreds of years. The very-high mass removal rate removes the envelope in about two to three years, which is about three times the Keplerian period when the companion enters the CEE at $R(12)=400 R_\odot$ and about 10 times the Keplerian orbit when it enters a common envelope at an earlier phase when $R(12)=200 R_\odot$. Here `12' stands for the mass of the RSG in solar units. 

We conduct some simulations on a thermal timescale because \cite{HiraiMandel2022} take the second stage of their proposed CEE formalism to be on a thermal timescale. The first stage in their proposed prescription is on a dynamical timescale, and so we also present  results for a mass removal on a dynamical timescale. We note that the spiralling-in heats the envelope even in envelope zones inner to the orbit of the companion. A spiralling-in on a dynamical timescale therefore, substantially increases the luminosity of the star. This would increase the convective zone for the same given envelope structure. This is one reason that we do not go to even shorter mass removal timescales. Another reason not do simulate shorter mass removal timescales is that in the inner zones of the envelope the spiralling-in process enters the self-regulated phase that proceeds on a timescale much longer than the local dynamical timescale (e.g., \citealt{GlanzPerets2021b}).   

We then follow the mass $M_{\rm RSG}$ and radius $R_{\rm RSG}$ of the star, and those of the boundary between the bottom of the envelope convective zone and the radiative zone above the core, $m_{\rm RC}$ and $R_{\rm RC}$, respectively. 

% =============================================
\section{Mass removal on a thermal timescale}
\label{sec:RadiativeZone}
% =============================================

We first present the results of high mass removal rate (much above the regular stellar wind) of $\dot M_{\rm CEE} = 0.01 M_\odot \yr^{-1}$ for a stellar model with an initial metalicity of $Z=0.01$ and when the RSG radius was $R(12)=400 R_\odot$, its core mass is $M_{\rm core} =2.5402 M_\odot$, and its luminosity is $L(12)=2.33\times 10^{4} L_\odot$, where `12' is the mass of the star in solar units. At that phase the core is a helium core and only hydrogen burns in a shell. 

\begin{figure}[h]
	\centering
%	\hspace*{-2cm} 
	% [trim=left bottom right top, clip]{file}
%	\hspace{1cm}
\includegraphics[trim=0.5cm 7.5cm 0.0cm 7.0cm ,clip, scale=0.45]{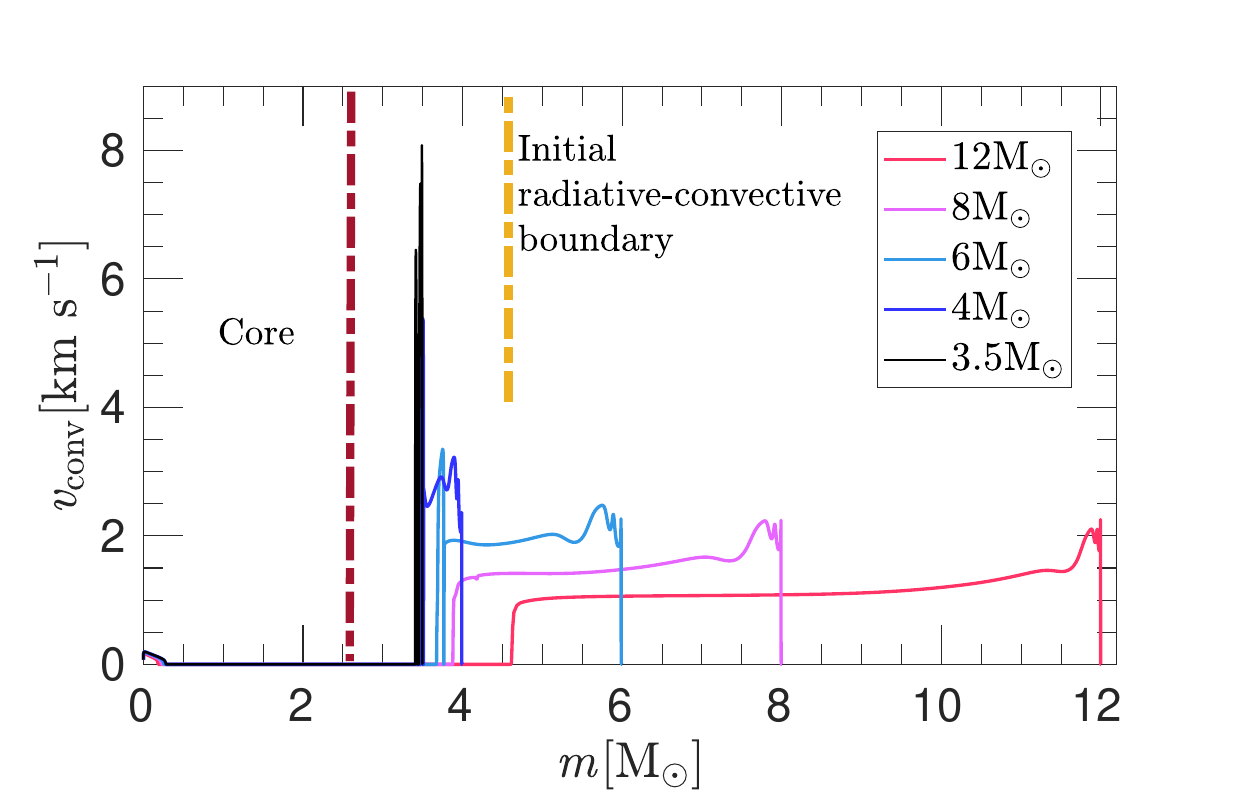}
\includegraphics[trim=0.5cm 7.5cm 0.0cm 7.0cm ,clip, scale=0.45]{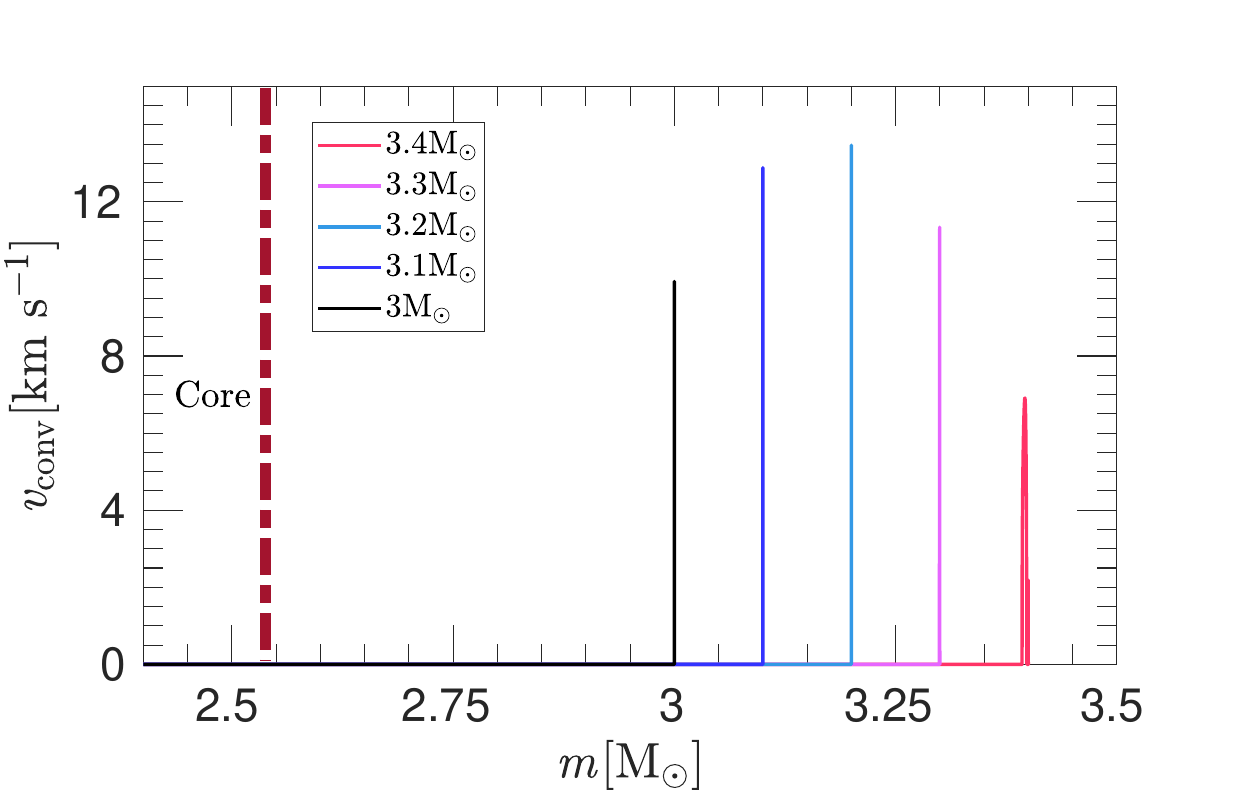}
\caption{The convective velocity as a function of mass coordinate at several times during mass removal that mimic a CEE on a thermal timescale of the envelope. Initial metalicity is $Z=0.01$. Mass removal rate is $\dot M_{\rm CEE} = 0.01 M_\odot \yr^{-1}$ so that the total evolution time that we present is 900 years. Upper panel: Profiles at early times. The vertical-dashed line represents the mass of the core, which has a negligible structural changes during the rapid mass removal. We note the decrease in the mass of the radiative zone between the core and the inner boundary of the extended envelope convective zone. Lower panel: Focusing on the collapse of the envelope to small radii as we continue to remove mass.
Insets show the total stellar mass $M_{\rm RSG}$ at each evolutionary point. 
Note the different scales of the axes in the two panels. 
}
\label{fig:VconVsMassCEE}
\end{figure}

In Fig. \ref{fig:VconVsMassCEE} we present the convective velocity as function of mass. The lines become thinner as we remove mass. The vertical dash line is the location of the outer boundary of the core. Where $v_{\rm conv} >0$ the envelope is convective. 
This figure clearly shows that the inner boundary of the envelope convective zone moves inward, namely, the mass in the radiative zone decreases. The initial radiative-convective boundary is at $m_{\rm RC} (12)=4.59 M_\odot$. The mass in the inner radiative zone of the envelope is $M_{\rm rad} (12) =m_{\rm RC}(12)- M_{\rm core} = 2.05 M_\odot$. By the time we remove the initial mass of the convective zone, namely when the stellar mass becomes $M=m_{\rm RC} (12)=4.59 M_\odot$, the convection did not disappear but rather the inner boundary of the envelope convective zone has penetrated deep into the initial radiative zone (in mass coordinate). 
%FFFFFFFFFFFFFFFFFFFFFFFFFFFFFFFFFFFFFFFFFFFFFFFFFF 
%FFFFFFFFFFFFFFFFFFFFFFFFFFFFFFFFFFFFFFFFFFFFFFFFFF

\begin{figure}
	\centering
%	\hspace*{-2cm} 
	% [trim=left bottom right top, clip]{file}
%	\hspace{1cm}
\includegraphics[trim=1.1cm 5.3cm 0.0cm 2.9cm ,clip, scale=0.46]{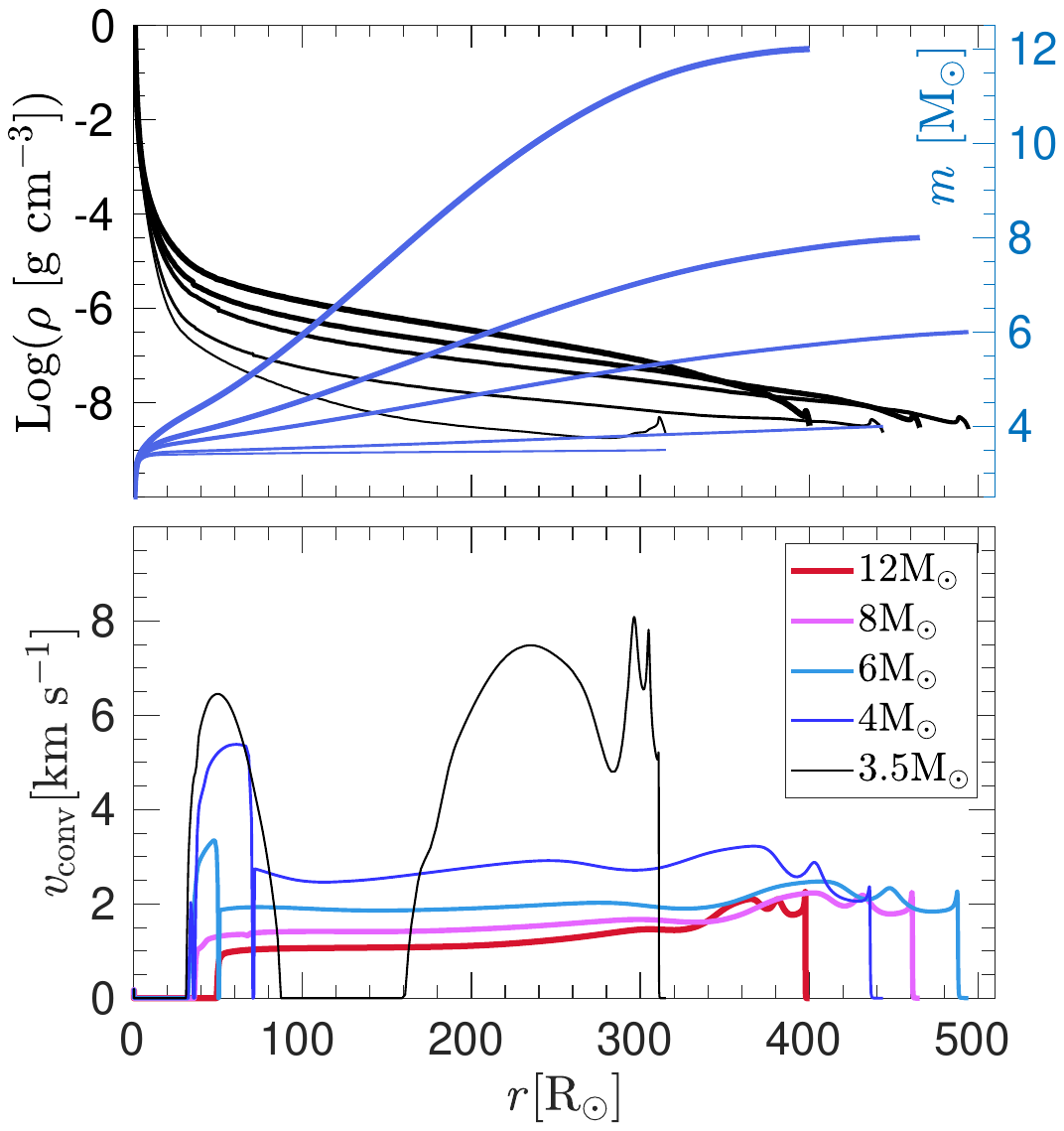} 
\caption{ Density $\rho(r)$ (black lines scale on the left of top panel), mass $m(r)$ (blue lines scale on the right of top panel), and the convective velocity $v_{\rm conv}(r)$ (coloured lines on lower panel) as function of radius coordinate at five evolutionary points as those in the upper panel of Fig. \ref{fig:VconVsMassCEE}. Inset shows the total stellar mass $M_{\rm RSG}$ at each evolutionary point. }
\label{fig:VconDenMassVsRCEE}
\end{figure}

In Fig. \ref{fig:VconDenMassVsRCEE} we present the relevant envelope properties as function of the radius. In the upper panel we present the mass and density as function of radius and in the lower panel the convective velocity. This figure clearly shows that the initial (when we start mass removal) mass coordinate of the radiative-convective boundary, $m_{\rm RC} (12)=4.59 M_\odot$, moves to larger radii as we remove mass. Mass shells that started in the radiative zone move out to become the inner region of the envelope convective zone. By the time we remove the entire initial envelope convective zone, when the stellar mass becomes  $M_{\rm RSG} \simeq 4.5 M_\odot$, a large fraction of the mass in the initial radiative zone has moved to larger radii and it is part of the envelope convective zone. 
The splitting of the convective zone when the envelope mass of the giant envelope becomes low is known to occur also in low mass stars (e.g., \citealt{SokerHarpaz1992}). 
%FFFFFFFFFFFFFFFFFFFFFFFFFFFFFFFFFFFFFFFFFFFFFFFFFF 
%FFFFFFFFFFFFFFFFFFFFFFFFFFFFFFFFFFFFFFFFFFFFFFFFFF

Above mass coordinate $m\simeq 3.5M_\odot$ the envelope has its ZAMS composition. Below that mass coordinate the envelope is helium rich.
At $m=3.5M_\odot$ the hydrogen fraction is $X=0.43$, decreasing more or less linearly with decreasing mass coordinate to $X=0$ at $m\simeq 2.3 M_\odot$.  The final envelope mass removal involves the helium-rich envelope. 

\begin{figure}
	\centering
%	\hspace*{-2cm} 
	% [trim=left bottom right top, clip]{file}
%	\hspace{1cm}
\includegraphics[trim=0.8cm 5.3cm 0.0cm 3.5cm ,clip, scale=0.46]{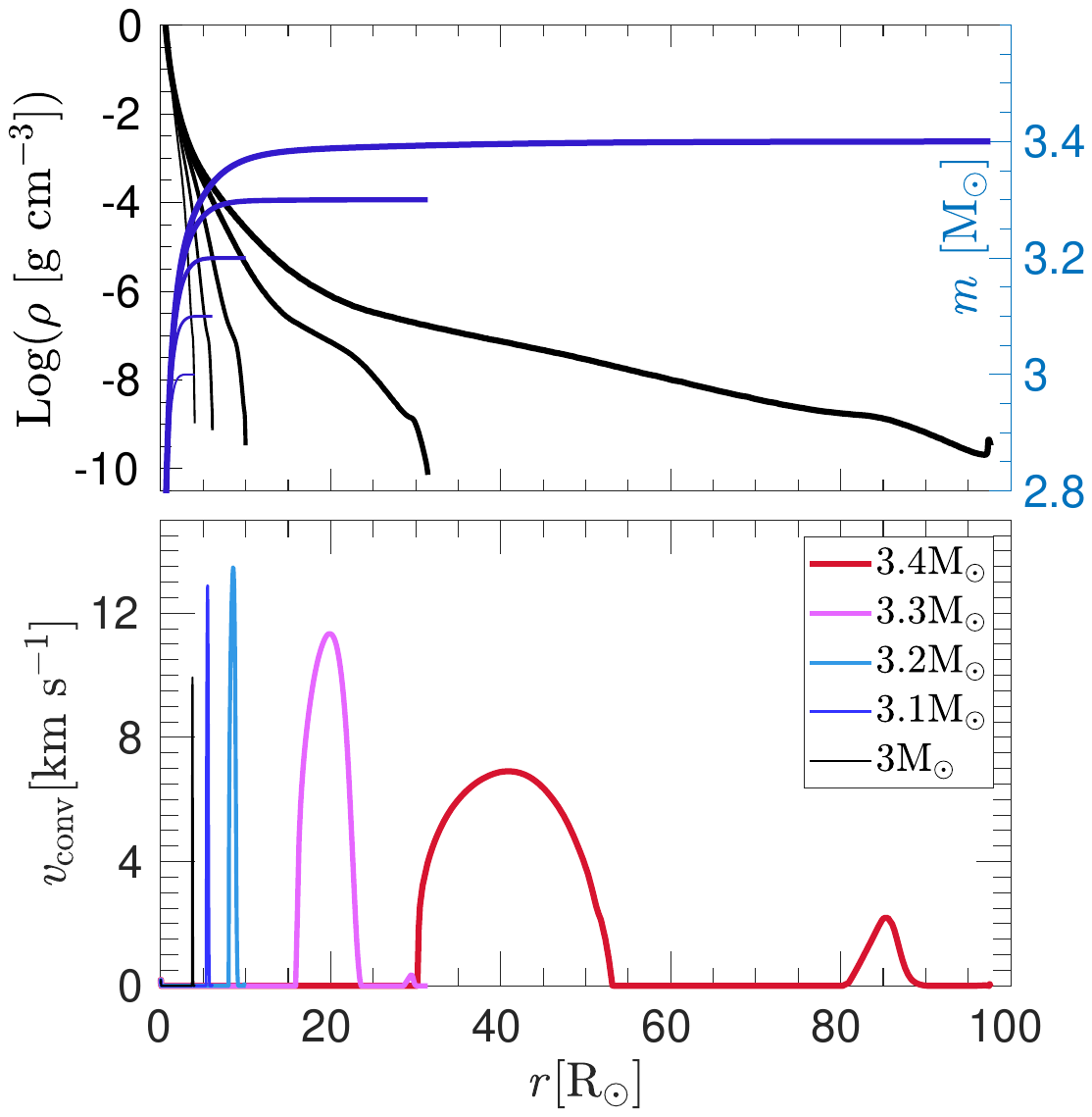} 
\caption{ Similar to Fig. \ref{fig:VconDenMassVsRCEE} but at later evolutionary times as in the lower panel of Fig. \ref{fig:VconVsMassCEE}.  
}
\label{fig:VconDenMassVsRFinal}
\end{figure}

In the lower panel of Fig. \ref{fig:VconVsMassCEE} we present the convective velocity as function of mass coordinate focusing on the final removal phase of our study (we stop before total envelope removal). Note that the mass axis of the lower panel of Fig. \ref{fig:VconVsMassCEE} starts at $m=2.5 M_\odot$. 
Fig. \ref{fig:VconDenMassVsRFinal} is similar to Fig. 
\ref{fig:VconDenMassVsRCEE} but at five late evolutionary points as in the lower panel of Fig. \ref{fig:VconVsMassCEE}. 
From the lower panel of Fig. \ref{fig:VconVsMassCEE} and from Fig.  \ref{fig:VconDenMassVsRFinal} we learn that the convective envelope zone practically disappears only when the envelope collapses to very small radii, much smaller than the initial radiative-convective boundary $R_{\rm RC}(12)$. %FFFFFFFFFFFFFFFFFFFFFFFFFFFFFFFFFFFFFFFFFFFFFFFFFF 
%FFFFFFFFFFFFFFFFFFFFFFFFFFFFFFFFFFFFFFFFFFFFFFFFFF

\begin{figure}
	\centering
%	\hspace*{-2cm} 
	% [trim=left bottom right top, clip]{file}
\includegraphics[trim=4.0cm 5.0cm 0.0cm 1.8cm ,clip, scale=0.35]{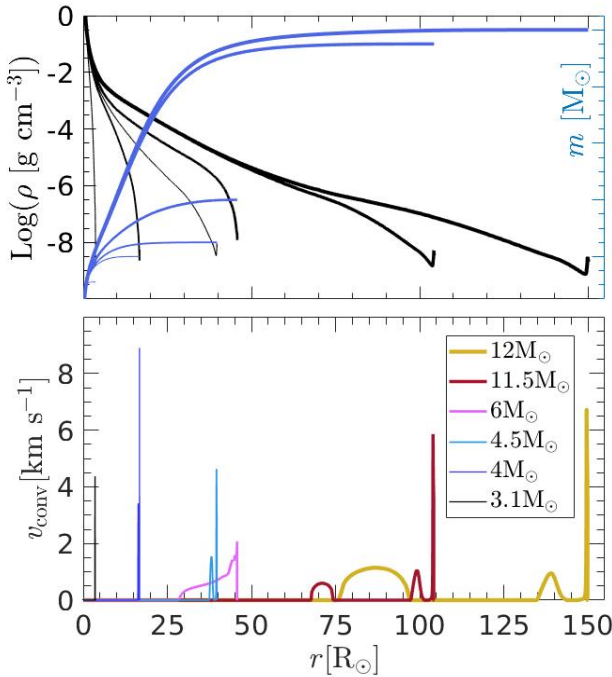} 
 \\ 
\caption{Similar to Fig. \ref{fig:VconDenMassVsRCEE} but when we start the high mass removal rate when the RSG radius is $R_{\rm RSG}(12)=150 R_\odot$. Note that some lines are at different RSG masses than in previous figures. 
}
\label{fig:VconDenMassVsRearly}
\end{figure}

In Fig. \ref{fig:VconDenMassVsRearly} we present the results for evolution that mimics the formation of a CEE at an earlier time when the RSG radius is $R_{\rm RSG}(12)=150 R_\odot$, its core mass is $M_{\rm core} = 2.5397 M_\odot$ and the luminosity is $L=1.77 \times 10^4 L_\odot$. We present the density, mass, and convective velocity at six evolutionary points. This case starts with very extended radiative zones in the envelope and  three convective zones. Despite the extended radiative zones the qualitative results for this case are the same as in the previous case of mass removal at a later evolutionary age (Figs. \ref{fig:VconDenMassVsRCEE} and \ref{fig:VconDenMassVsRFinal}). There is a persistent envelope convective zone (or 2-3 zones) until the envelope rapidly shrinks (collapses). Namely, the envelope convective zone deepens into the initial envelope radiative zone (in mass coordinate). 
%FFFFFFFFFFFFFFFFFFFFFFFFFFFFFFFFFFFFFFFFFFFFFFFFFF 
%FFFFFFFFFFFFFFFFFFFFFFFFFFFFFFFFFFFFFFFFFFFFFFFFFF

We repeat the same procedure of high mass removal rate at $R_{\rm RSG}(12)=150 R_\odot$ and  $R_{\rm RSG}(12)=400 R_\odot$ but for a stellar model with an initial metallicity of $Z=0.02$. We find the results to be very similar to the $Z=0.01$ cases that we discussed above. 

\begin{figure}
	\centering
%	\hspace*{-2cm} 
	% [trim=left bottom right top, clip]{file}
%	\hspace{1cm}
\includegraphics[trim=1.2cm 5.3cm 0.0cm 2.9cm ,clip, scale=0.46]{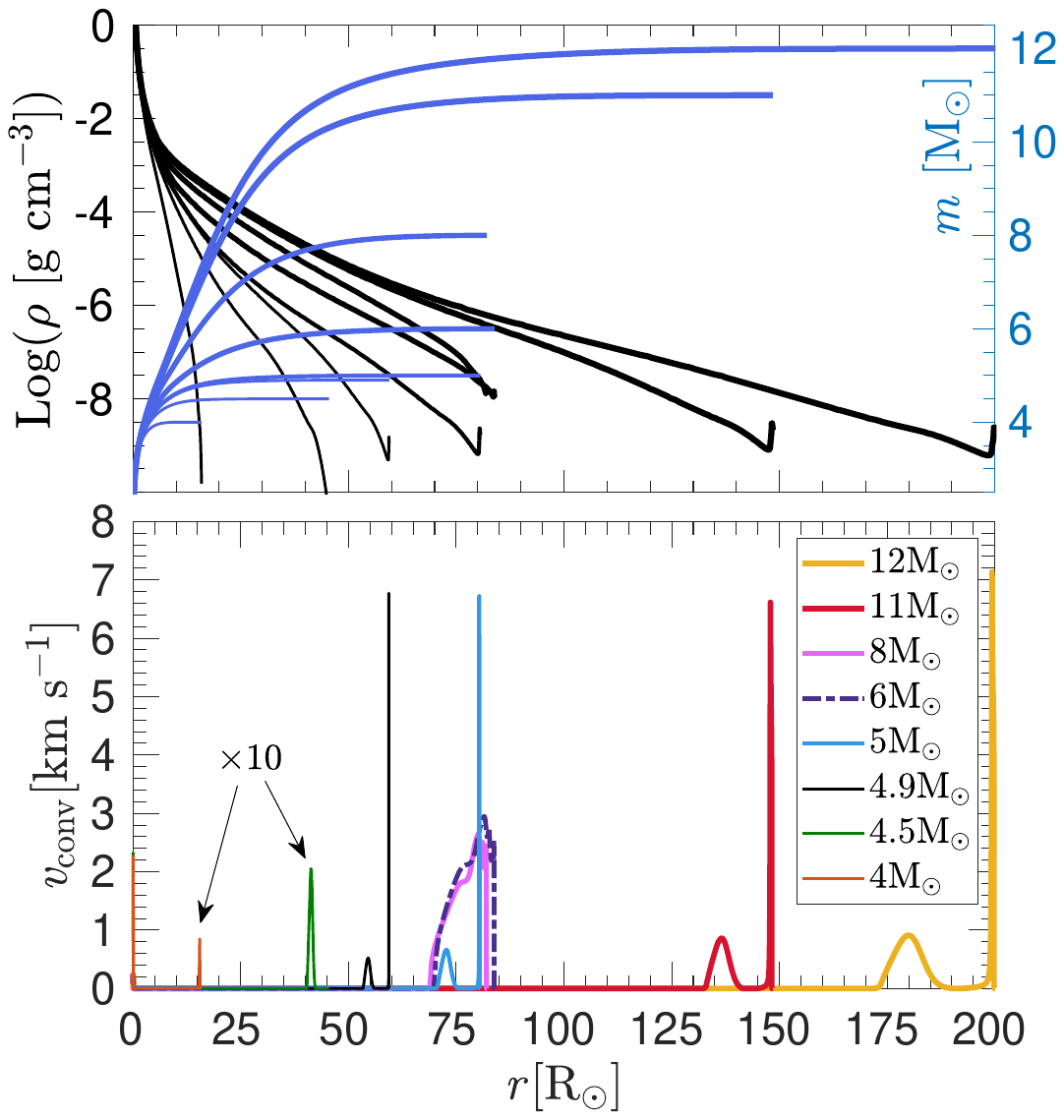} 
\caption{Similar to Fig. \ref{fig:VconDenMassVsRCEE} but at eight evolutionary times and for a stellar model with a lower initial metalicity of $Z=0.001$. We start the high-rate mass removal process when the RSG radius is $R_{\rm RSG}(12)=200 R_\odot$. Note that the last two (on the left) lines of the convective velocity in the lower panel are multiplied by 10.  
}
\label{fig:VconDenMassVsRLowMetal}
\end{figure}

In Fig. \ref{fig:VconDenMassVsRLowMetal} we present the envelope evolution during the high mass removal rate ($\dot M_{\rm CEE}=0.01 M_\odot \yr^{-1}$) for a stellar model with a much lower initial metalicity of $Z=0.001$. We start to remove mass when the RSG radius is $R(12)=200 R_\odot$. At that time the core mass is $M_{\rm core} =2.636 M_\odot$ and the luminosity is $L(12)=2.91 \times 10^4 L_\odot$.
In this case the initial envelope radiative zones covers most of the envelope. The mass in the two convective zones is very small. We clearly see that the inner boundary of the convective zone(s) penetrate deep (in mass coordinate) into the initial radiative zones. The convective zone disappears (or almost disappear) only when the RSG envelope substantially shrinks (collapses).     
%FFFFFFFFFFFFFFFFFFFFFFFFFFFFFFFFFFFFFFFFFFFFFFFFFF 
%FFFFFFFFFFFFFFFFFFFFFFFFFFFFFFFFFFFFFFFFFFFFFFFFFF

% =============================================
\section{Mass removal on a dynamical timescale}
\label{sec:RadiativeZonedynamic}
% =============================================

In this section we present the results of very-high mass removal rate of $\dot M_{\rm CEE,d} = 3 M_\odot \yr^{-1}$, i.e., on a dynamical timescale. 

\begin{figure}[h]
	\centering
%	\hspace*{-2cm} 
	% [trim=left bottom right top, clip]{file}
%	\hspace{1cm}
\includegraphics[trim=0.5cm 7.5cm 0.0cm 7.0cm ,clip, scale=0.45]{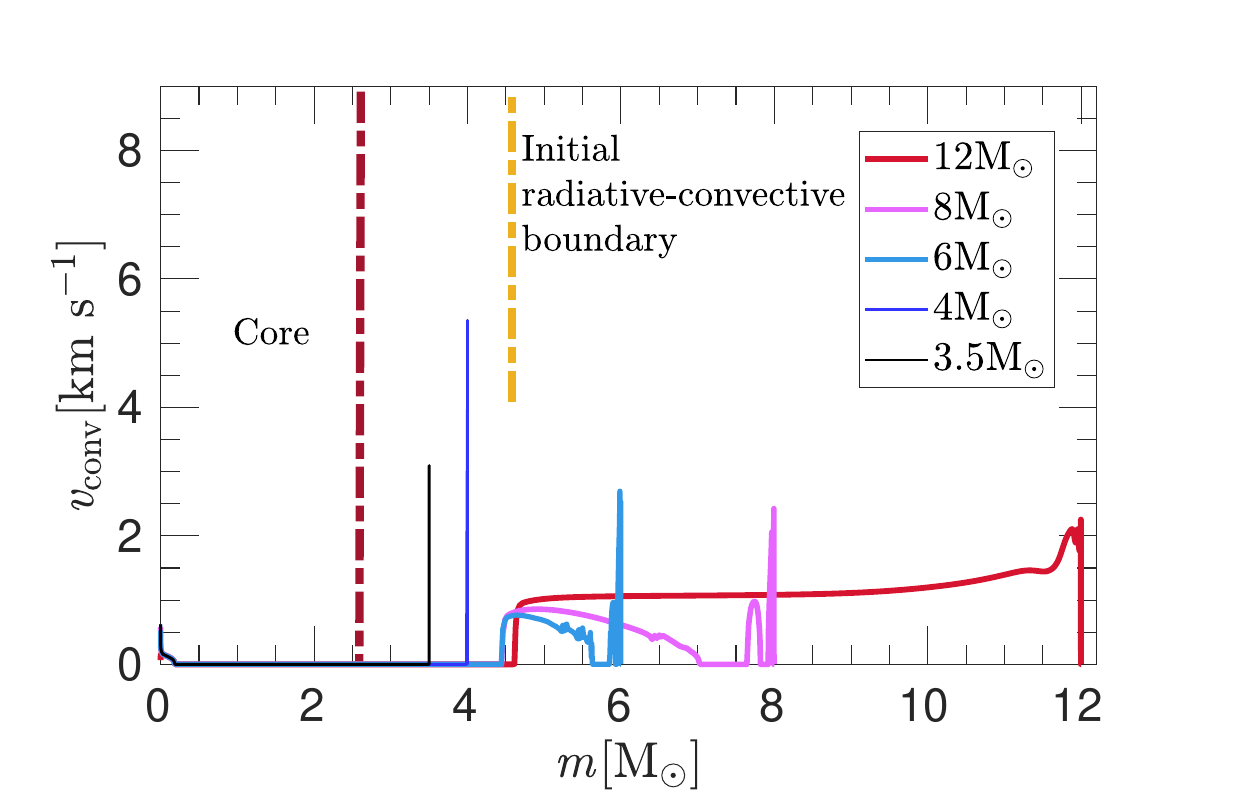}
\includegraphics[trim=0.5cm 7.5cm 0.0cm 7.0cm ,clip, scale=0.45]{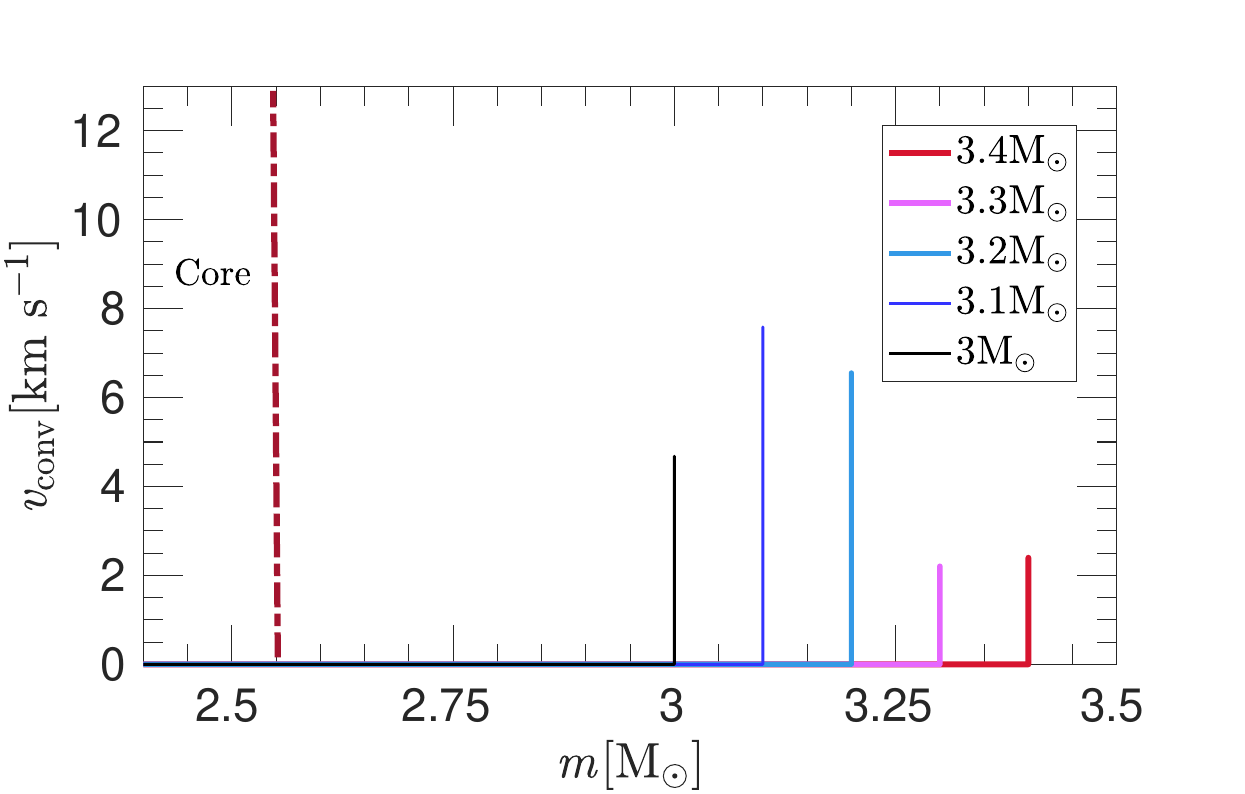}
\caption{Similar to Fig. \ref{fig:VconVsMassCEE} but for the very-high mass removal rate of $\dot M_{\rm CEE,d} = 3 M_\odot \yr^{-1}$.  
}
\label{fig:VconVsMassCEEdynamic}
\end{figure}

\begin{figure}
	\centering
%	\hspace*{-2cm} 
	% [trim=left bottom right top, clip]{file}
%	\hspace{1cm}
\includegraphics[trim=1.1cm 5.3cm 0.0cm 2.9cm ,clip, scale=0.46]{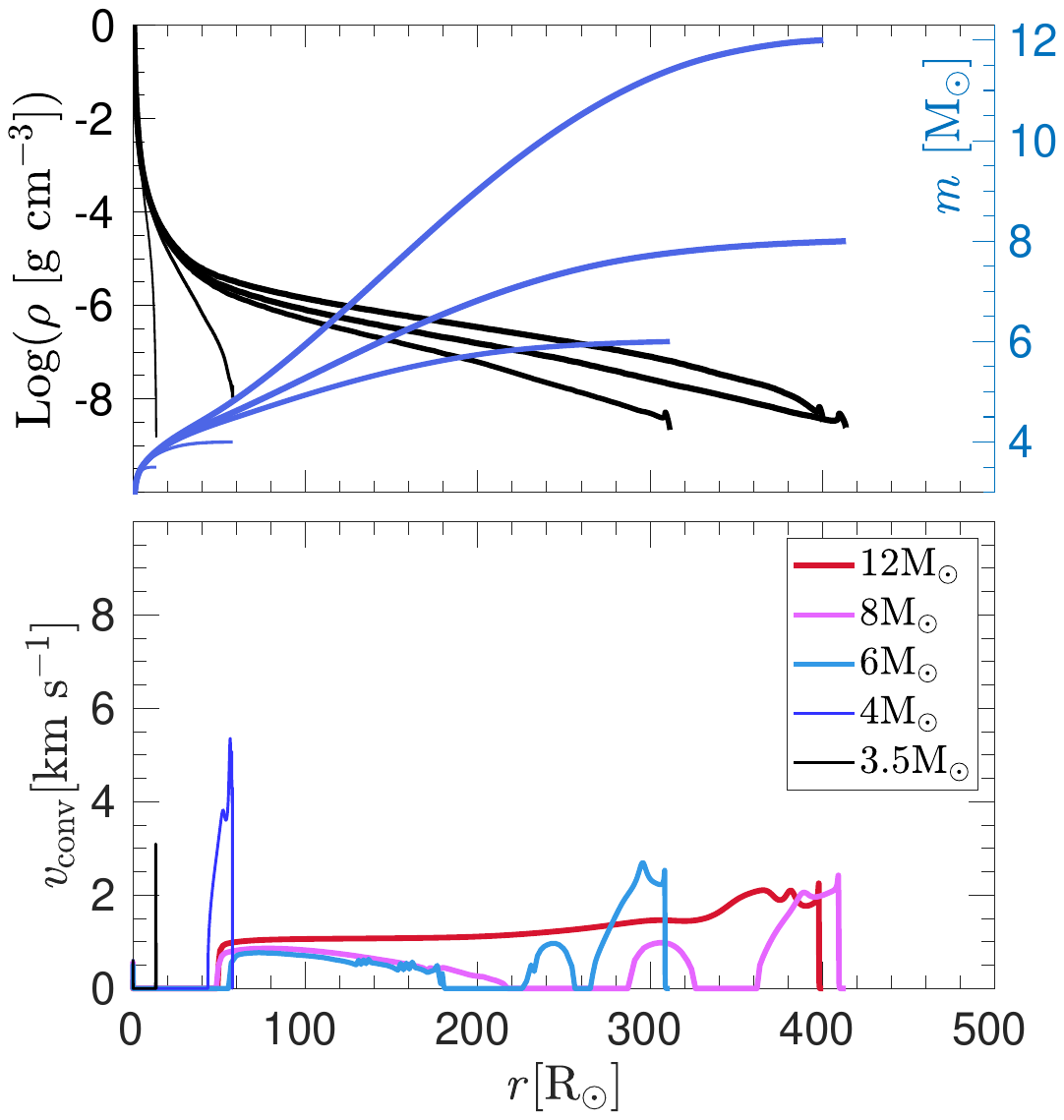} 
\caption{Similar to Fig. \ref{fig:VconDenMassVsRCEE} but for the very-high mass removal rate of $\dot M_{\rm CEE,d} = 3 M_\odot \yr^{-1}$.  }
\label{fig:VconDenMassVsRCEEdynamic}
\end{figure}

\begin{figure}
	\centering
%	\hspace*{-2cm} 
	% [trim=left bottom right top, clip]{file}
%	\hspace{1cm}
\includegraphics[trim=0.8cm 5.3cm 0.0cm 3.5cm ,clip, scale=0.46]{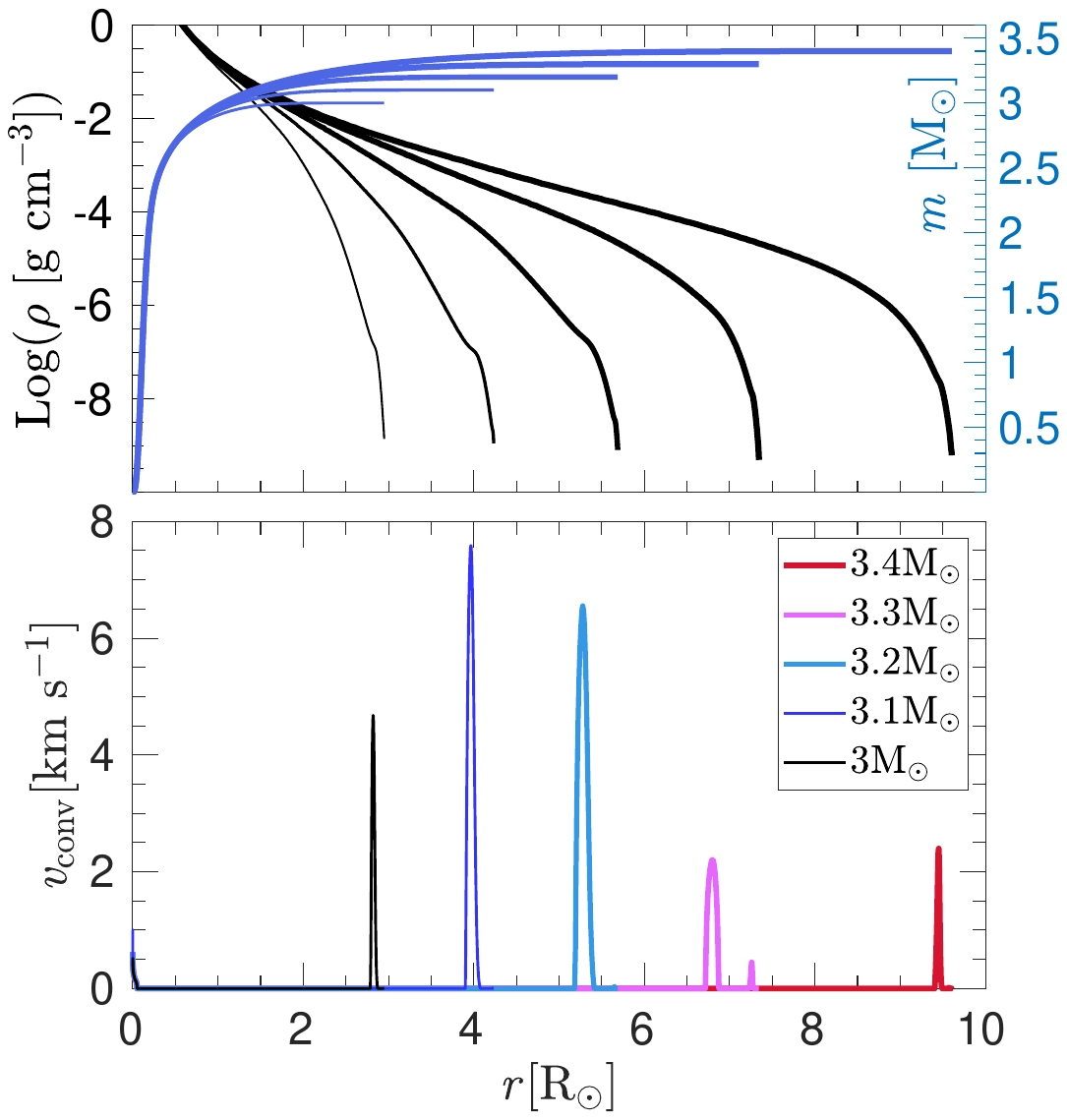} 
\caption{ Similar to Fig. \ref{fig:VconDenMassVsRCEEdynamic} but at later evolutionary points as in the lower panel of Fig. \ref{fig:VconVsMassCEEdynamic}. Namely, similar to Fig. \ref{fig:VconDenMassVsRFinal} but for the very-high mass removal rate; 
note the different scales of the axes. 
}
\label{fig:VconDenMassVsRFinaldynamic}
\end{figure}

In Figs. \ref{fig:VconVsMassCEEdynamic} - \ref{fig:VconDenMassVsRFinaldynamic} we present the results for the same model and the same starting point as in Figs. \ref{fig:VconVsMassCEE} - \ref{fig:VconDenMassVsRFinal}, respectively. 
Namely, for a stellar model with an initial metalicity of $Z=0.01$ and when the RSG radius was $R(12)=400 R_\odot$. 
%FFFFFFFFFFFFFFFFFFFFFFFFFFFFFFFFFFFFFFFFFFFFFFFFFF 
%FFFFFFFFFFFFFFFFFFFFFFFFFFFFFFFFFFFFFFFFFFFFFFFFFF
%FFFFFFFFFFFFFFFFFFFFFFFFFFFFFFFFFFFFFFFFFFFFFFFFFF 
%FFFFFFFFFFFFFFFFFFFFFFFFFFFFFFFFFFFFFFFFFFFFFFFFFF
%FFFFFFFFFFFFFFFFFFFFFFFFFFFFFFFFFFFFFFFFFFFFFFFFFF 
%FFFFFFFFFFFFFFFFFFFFFFFFFFFFFFFFFFFFFFFFFFFFFFFFFF

\begin{figure}
	\centering
%	\hspace*{-2cm} 
	% [trim=left bottom right top, clip]{file}
%	\hspace{1cm}
\includegraphics[trim=1.2cm 5.3cm 0.0cm 2.9cm ,clip, scale=0.46]{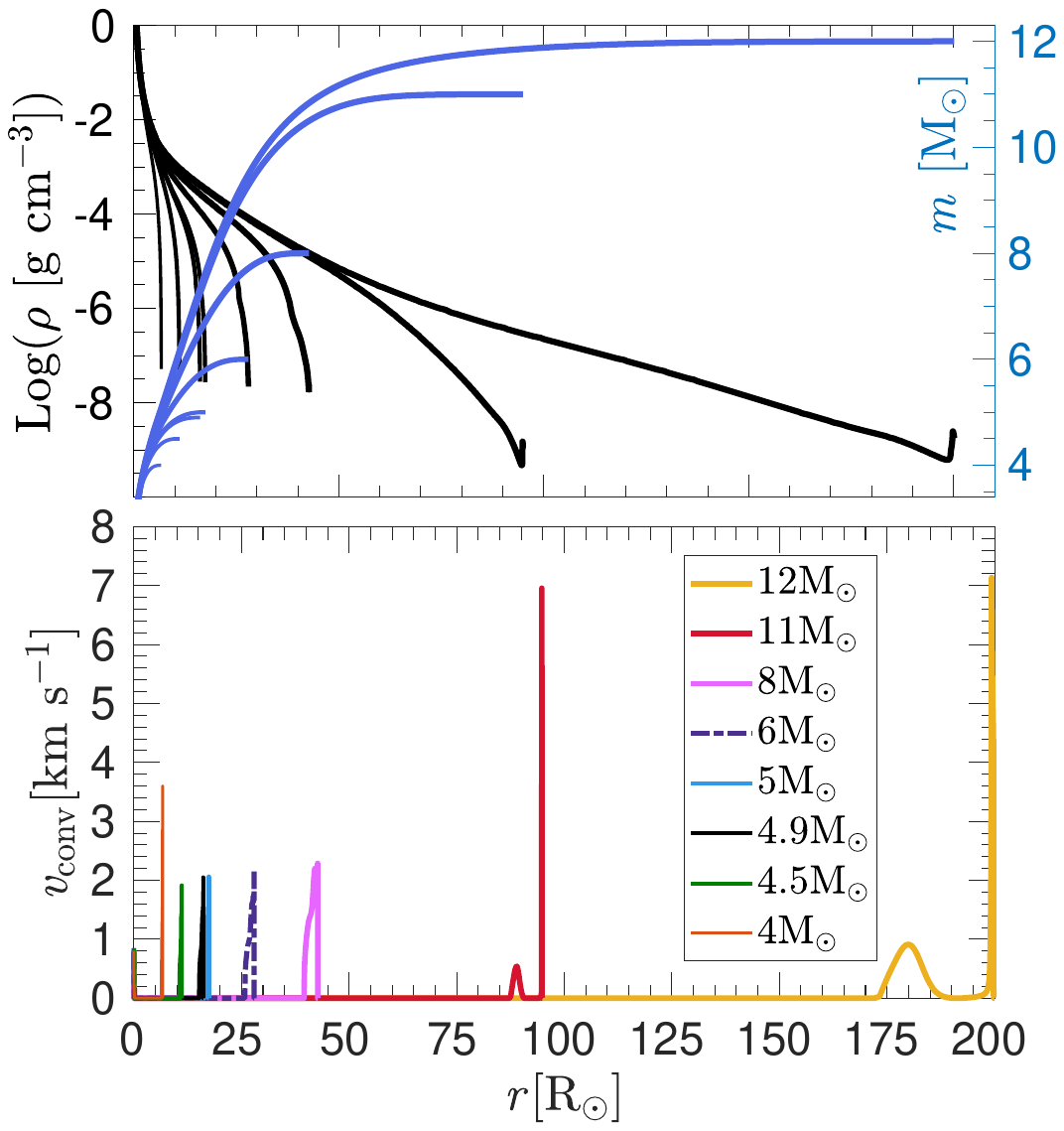} 
\caption{Similar to Fig. \ref{fig:VconDenMassVsRLowMetal}, i.e., for an initial metalicity of $Z=0.001$, but when the mass removal rate is $\dot M_{\rm CEE,d} = 3 M_\odot \yr^{-1}$. The mass removal on a dynamical timescale leads to a faster contraction of the convective zones. However,  the behavior is qualitatively similar to that in the case of mass removal on a thermal timescale.  
}
\label{fig:VconDenMassVsRLowMetaldynamic}
\end{figure}

In Fig. \ref{fig:VconDenMassVsRLowMetaldynamic} we present the results for the low-metalicity model, $Z=0.001$, but for the very-high mass removal rate of $\dot M_{\rm CEE,d} = 3 M_\odot \yr^{-1}$ (on a dynamical timescale). We start to remove mass at the same evolutionary point as in the simulation we present in Fig. \ref{fig:VconDenMassVsRLowMetal}.  
%FFFFFFFFFFFFFFFFFFFFFFFFFFFFFFFFFFFFFFFFFFFFFFFFFF 
%FFFFFFFFFFFFFFFFFFFFFFFFFFFFFFFFFFFFFFFFFFFFFFFFFF

Comparing Figs. \ref{fig:VconVsMassCEEdynamic} - \ref{fig:VconDenMassVsRFinaldynamic} to Figs. \ref{fig:VconVsMassCEE} - \ref{fig:VconDenMassVsRFinal} and Fig. \ref{fig:VconDenMassVsRLowMetaldynamic} to Fig. \ref{fig:VconDenMassVsRLowMetal} we find that the behavior of the envelope convective zones under mass removal on a dynamical timescale is qualitatively similar to the behavior under mass removal on a thermal timescale. Namely, the initial radiative zone when we start to remove mass that mimics a CEE does not maintain its structure. Rather the convective zone penetrates into the initial radiative zone and the radiative zone contracts. 

The main quantitative difference is that the envelopes contracts faster during the dynamical mass removal. 
The more rapid contraction in the simulations of dynamical mass removal (very-high mass removal rate) is expected as radiative layers tend to shrink under mass removal on dynamical timescales (e.g., \citealt{Ivanova2011}). 

We compare our results to the \textsc{mesa} simulation by \cite{Fragosetal2019}. They use \textsc{mesa} to simulate the spiralling-in of a neutron star of $1.4 M_\odot$ into the envelope of a $12 M_\odot$ RSG. They calculate the orbital separation of the neutron star inside the envelope but did not allow for mass loss. Namely, their envelope expanded to $\simeq 10^4 R_\odot$ but no mass loss occurred. This non-mass loss simulation changes somewhat the properties of the envelope near the orbit of the companion. We can see this in figure 1 of \cite{Fragosetal2019} by following the convective zone. From about $t \simeq 1.5 \yr$ to $t\simeq 7 \yr$ in their simulation the inner boundary of the convective zone is moving outward in radius. At $t \simeq 7 \yr$, after the envelope further expands, a convective zone appears just above the orbit of the neutron star. It seems that if a real mass ejection process was allowed the convective zone near the neutron star orbit would have been there along the entire evolution. The new convective zone appears inner to the initial convective zone both in mass and radial coordinates. Our results show a similar behavior.  

The orbital separation of the neutron star from the center of the RSG in the simulation by \cite{Fragosetal2019} shows the three phases of the CEE: slow decrease as the orbit-spin synchronisation is lost, a rapid plunge-in phase, and a slow spiralling-in during the self-regulated phase. At $t=10 \yr$ in their simulation (last time they present results for the orbit) the orbital separation is about $12 R_\odot$ which is half the initial inner boundary of the envelope convective zone. Namely, the neutron star companion spiralled-in deep into the initial radiative zone in radial coordinate. 
Eventually the envelope contract in their simulation, as we also find here, and can shrink inner to the orbit of the neutron star. This is expected in all CEE evolution where the companion survives. Then a new phase of binary evolution starts where other processes might take place, like a circumbinary disk and the launching of jets. 

There are no contradictions between our claim and results and the results of  \cite{Fragosetal2019}. We do not claim that the spiralling-in does not slows down at a small orbital separation, sometimes close to the initial boundary between the radiative zone and convective zone. Our claim is only that the initial location of this boundary cannot be used to predict the final orbital separation.

% =============================================
\section{Implications}
\label{sec:Implications}
% =============================================

The results of sections \ref{sec:RadiativeZone} and \ref{sec:RadiativeZonedynamic}
as we present in Figs. \ref{fig:VconVsMassCEE} to \ref{fig:VconDenMassVsRLowMetaldynamic} show that by the time the CEE removes the initial convective zone of the envelope the inner boundary of the envelope convective zone has moved deep into the envelope, now including a large fraction of the initial inner radiative zone of the envelope (above the core). 
The convective zone practically disappears (or becomes very thin) only when the stellar radius decreases to be much lower than the initial boundary of the radiative and convective zones, $R_{\rm RSG} \la 0.1 R_{\rm RC}(12)$. 
By the time the convective envelope becomes very thin the inner radiative zone of the envelope contains much less mass than its initial mass. 
The immediate conclusion is that one cannot split the CEE to two stages, before and after the removal of the initial convective zone of the RSG envelope. 

This late disappearance of the envelope convective zone is known to occur also in low mass stars that evolve from the AGB to the post-AGB phase (e.g., \citealt{Soker1992}).  

We argue, therefore, that the proposal by \cite{HiraiMandel2022} of a two-stage formalism for CEE of massive stars is problematic. They propose to use the $\alpha$-formalism while the compact companion spirals-in inside the initial convective zone. They then assume that the system enters a stable mass transfer from the initial radiative zone to the companion.
However, our finding that the inner boundary of the envelope convective zone moves deep into the initial radiative zone makes this formalism questionable.

During the CEE the spiralling-in companion deposits energy to the envelope. In this study we did not add such an energy to the envelope because we focus our study on  the second stage as proposed by \cite{HiraiMandel2022}. That stage proceeds on a thermal timescale and therefore the orbital power is lower than the RSG stellar luminosity.

% =============================================
\section{Summary}
\label{sec:Summary}
% =============================================

We used \textsc{mesa} (section \ref{sec:Numerics}) to evolve RSG models with an initial mass of $M_{\rm ZAMS}=12 M_\odot$ through a rapid mass removal phase to mimic a CEE. We studied the removal of mass during hydrogen-shell burning when the RSG star has a helium core on either a thermal timescale (section \ref{sec:RadiativeZone}) or on a dynamical timescale (section \ref{sec:RadiativeZonedynamic}).
We present the evolution of the envelope convective zones during the high mass removal rate of $\dot M_{\rm CEE} = 0.01 M_\odot \yr^{-1}$ for a model with an initial metalicity of $Z=0.01$ and for which we start to remove mass when the RSG radius was $R_{\rm RSG}(12)=400 R_\odot$ (Figs.  \ref{fig:VconVsMassCEE}-\ref{fig:VconDenMassVsRFinal}), when its radius was $R_{\rm RSG}(12)=150 R_\odot$ (Fig. \ref{fig:VconDenMassVsRearly}), and for a stellar model with a much lower initial metalicity of $Z=0.001$ and for which we start to remove mass when the RSG radius was $R_{\rm RSG}(12)=200 R_\odot$ (Fig. \ref{fig:VconDenMassVsRLowMetal}). 
We repeat the case of Figs. \ref{fig:VconVsMassCEE}-\ref{fig:VconDenMassVsRFinal} and the case of Fig. \ref{fig:VconDenMassVsRLowMetal} but with a very-high mass removal rate of $\dot M_{\rm CEE,d} = 3 M_\odot \yr^{-1}$, i.e., on a dynamical timescale, and present the results in Figs. \ref{fig:VconVsMassCEEdynamic}-\ref{fig:VconDenMassVsRFinaldynamic} and Fig. \ref{fig:VconDenMassVsRLowMetaldynamic}, respectively.  
 
Our main result is that the inner boundary of the envelope convective zone moves deeper and deeper into the initial radiative zone as we remove mass. The convective zone practically disappears only after the entire RSG envelope shrinks (collapses) to a radius of $R_{\rm RSG} \la 0.1 R_{\rm RC}(12)$, namely, by about an order of magnitude smaller radius that the initial radiative-convective boundary. 

Our result implies that one cannot split the CEE into two stages as \cite{HiraiMandel2022} suggest: a stage when the companion spirals-in inside a convective envelope and a later stage of slow evolution when it orbits the initial radiative part of the envelope. Specifically, \cite{HiraiMandel2022} formalism requires the envelope radiative zone to expand, while we find that it contracts both in radius and mass. We find no ground for the two-stage CEE formalism that \cite{HiraiMandel2022} proposed.  

It is true that at the end of the CEE the small low-mass envelope of the giant is practically radiative and that it eventually becomes smaller than the orbital separation of surviving companions. However, this takes place much later than what \cite{HiraiMandel2022} assume. As well, by that time, i.e., when the orbital separation is $a \la 10 R_\odot$, other processes might play significant roles (e.g., \citealt{Soker2017FinalCEE}). One processes is the possible formation of a circumbinary disk at the termination of the CEE (\citealt{KashiSoker2011}). Another processes might occur for main sequence companions. A main sequence companion accretes mass during the CEE and might swell.
As it approaches the core of the RSG it might lose mass via a Roche lobe overflow. The removal of this extra mass requires energy that comes from the orbital energy. Therefore, this processes causes further spiralling-in. 

It is possible that some systems exit the CEE phase in a grazing envelope evolution. Namely, the companion orbits on the surface of the, now small, giant accretes mass and launches jets \citep{Soker2017FinalCEE}. This stage can be long, as in the scenario of \cite{HiraiMandel2022}. However, we do not think this is related to the initial envelope radiative zone. As well, the companion might accrete from a circumbinary disk and launch jets (e.g., \citealt{Soker2014}) 

Yet, there is another process that might take place at the end of the CEE. As the common envelope collapses it can spin-up due to the decreasing moment of inertia and further spiralling-in of the companion to the degree that two opposites funnels are open along the polar axis, leading to jets (e.g., \citealt{Soker1992, Zouetal2020}). 

Finally, there is a possibility of two phases CEE as \cite{SokerBear2023} suggest in a recent study. In the first phase the companion removes the entire hydrogen-rich envelope, such that the hydrogen-poor core of the giant shrinks inside the orbit of the companion. Later, the CO core mass increases, the core contracts, and the  helium-rich layer expands and engulfs the companion to resume a second CEE phase. 

Following our results we hold the view that the $\alpha$-formalism is yet the best phenomenological formalism for population synthesis that involve CEE. Despite its simplicity, it is flexible. For example, one can take $\alpha_{\rm Ce} >1$ as some population synthesis studies do (e.g., \citealt{Garciaetal2021, Zevinetal2021, BroekgaardenBerger2021, Grichener2023}) because other sources of energy to the orbital energy exist, in particular jets that the companion can launch. 

We expect that the CEE does not have a constant value of $\alpha_{\rm CE}$ during the spiralling-in process of a given system. In some parts of the envelope jets are more powerful, like at the beginning and at the end of the CEE  (e.g., \citealt{Soker2017FinalCEE}). On the other hand, convection is more likely to transport orbital energy out when the companion is in the outer regions of the envelope (e.g., \citealt{WilsonNordhaus2022}). At the final CEE phase the gravity of the core might remove some or all of the mass that the companion accreted at earlier CEE phases. This will come on the expense of the orbital energy, further reducing the orbital separation. 

Our point is that the $\alpha$-formalism can phenomenologically accommodate the contribution of these energy sources and sinks, and at present it is the best CEE formalism for population synthesis codes.

%===========================================
\section*{Acknowledgements}
%===========================================
We thank Matthias Kruckow and an anonymous referee for useful comments. 
This research was supported by a grant from the Israel Science Foundation (769/20). 
% =======================

%%%%%%%%%%%%%%%%%%%%%%%%%%%
%\section*{Data availability}
%The data underlying this article will be shared on reasonable request to the corresponding author. 
%%%%%%%%%%%%%%%%%%%%%%%%%%%

\label{lastpage}
\end{document}